\documentclass[journal,10pt]{IEEEtran}
\usepackage[T1]{fontenc}
\usepackage{amsmath,amssymb,amsthm,graphicx,booktabs,tabularx,array,cite,url,microtype,enumitem}
\usepackage[hidelinks]{hyperref}
\usepackage{placeins,balance}

\newcommand{\etal}{\textit{et al.}}
\author{Isarar Siddique, Raiyan Siddique, Arshi Niyaz, Jamyang Ugyen Tshomo, and Atish Raza%
\thanks{All authors are with Convolity AI Lab (e-mails: isarar@convolity.com; raiyan@convolity.com; arshi@convolity.com; jamyang@convolity.com; atish@convolity.com). Corresponding author: Isarar Siddique.}}
\title{Evidence Field for Histopathology: Cross-Centre Evaluation and Replayable Spatial Decisions}
\begin{document}
\maketitle
\raggedbottom
\begin{abstract}
Histopathology classifiers must transfer across acquisition sites while exposing evidence that can be checked independently. Evidence Field couples learned image measurements and spatial classification with a native covering-program record. We evaluate this architecture on all 455,954 released CAMELYON17-WILDS patches, using patient-separated source partitions and two held-out hospital centres. Eighteen from-scratch fits compare four Evidence Field objectives with residual convolutional and compact vision-transformer classifiers. Checkpoints, temperatures and review thresholds are selected using source validation. The source-selected paired-cross-entropy ensemble achieves areas under the receiver operating characteristic curve of 0.9356 and 0.9453, with accuracies of 84.46\% and 86.77\% at the two held-out centres. Corresponding convolutional values are 0.7352 and 0.7656, and transformer values are 0.8502 and 0.7924. Evidence Field has 10,550 trainable parameters per member and additionally returns selected covers, signed region and interaction terms, contextual dependencies and independently reconstructible predictions. Patient-cluster intervals, calibration, source-fixed review policies and recorded training resources characterize the results. A 340-image, three-member audit reconstructs probability mixtures and 6,120 class programs under three rigid views. The evaluation supports cross-centre discrimination and reproducible computation for these implemented models; unequal family tuning, previously examined public targets and limited patient groups constrain broader interpretation.
\end{abstract}
\section{Introduction}
Histopathology transfer requires defined targets, independent development patients and held-out institutions~\cite{camelyon,camelyon16,campanella,clam}. A classifier's probability leaves its contributing measurements and decision path implicit. EF makes that computation explicit: a legal spatial-covering program selects the same signed contributions that generate its logits.
\begin{figure*}[!t]\centering
\includegraphics[width=\linewidth]{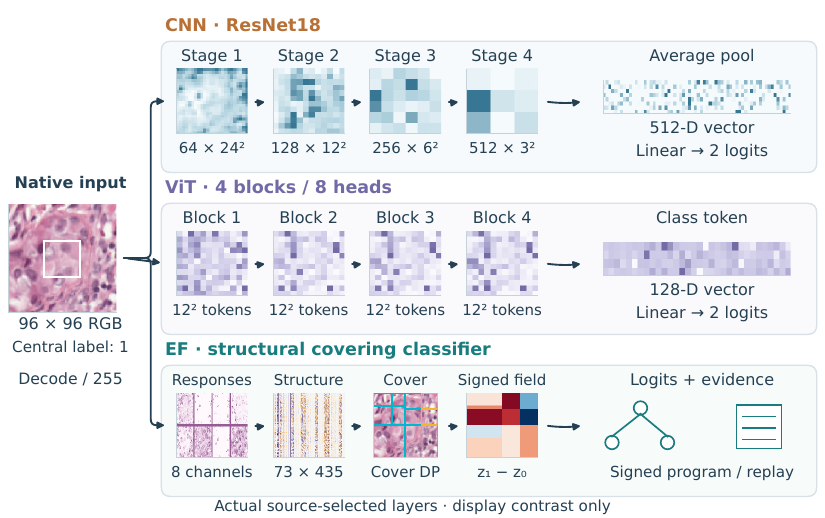}
\caption{Actual source-selected layer views on one training image. After its stride-two stem, CNN's four residual stages yield channel-RMS maps and a pooled 512-D vector. ViT embeds 8\(\times\)8-pixel patches into 144 tokens; four token-RMS maps lead to a 128-D class token. Both linear heads yield two logits. EF learns eight response channels, 435-D region descriptors and a covering program, returning logits with signed supports and a replay record (glyph). Vector reshaping and contrast stretching are display-only; the white square marks the label window.}
\label{fig:flow}\end{figure*}
\noindent Our contribution is a concrete test of native classification and replayable spatial computation. The comparison holds source images, epochs and seeds fixed, reports all four EF objectives, and separates ranking, decisions, confidence and evidence. The study tests the implemented patch classifiers and their evidence records.
\subsection{Nearest mechanisms} Radial units~\cite{moody} and Fourier features~\cite{rahimi}
precede the encoder; spatial pyramids~\cite{spatialpyramid} precede multiscale
aggregation, while semiring inference~\cite{goodman} and rule ensembles~\cite{ruleensemble}
precede structured reductions and interpretable terms. EF's proposed coupling
binds referenced signatures and cross-scale scores to one-fold selected covers
and independently reconstructible optima. The accompanying technical
manuscript~\cite{efcore} defines the core; this study evaluates its cross-centre
classification, calibration, cost and native evidence. The related architecture manuscript contains the trained mechanism and class-centred-loss comparisons. This application manuscript reports the original source-selected classifiers, their hospital transfer and native outputs; the shared experiments are identified rather than counted as independent studies.

\section{Clinical and computational context}
\subsection{Targets and scale} Tumour detection, molecular prediction and outcomes are different endpoints~\cite{coudray,kather_msi,pan_cancer,echle,kather_crc}; prostate studies emphasize independent reference standards~\cite{nagpal,bulten,panda}. Slide aggregation and origin prediction learn contextual representations~\cite{campanella,clam,toad}. Foundation models use other pretraining scales and supervision~\cite{path_foundation,path_language,virchow,chief,gigapath}; they are outside this from-scratch comparison. Native 96\(\times\)96 lymph-node patches also differ from rescaled benchmarks~\cite{medmnist}.
\subsection{Acquisition and evidence} Stain methods address optical variability~\cite{tellez,vahadane,ciga}; nucleus and multi-rater models supply biological localization~\cite{nuclei,hovernet,nucls}. EF's supports are computational regions. Numerical reconstruction~\cite{pyradiomics,ibsi} and hospital-shift auditing~\cite{zech,degrave} are distinct from clinical screening and referral validation~\cite{mckinney,defauw}.
\section{Data source, label definition and quality}
\subsection{Origin and access} CAMELYON17 contains H\&E lymph-node slides from five Dutch centres~\cite{camelyon}. Its \textbf{CAMELYON17-WILDS} derivative~\cite{wilds} releases 455,954 native 96\(\times\)96 RGB patches. We retained 21 Parquet files from the \href{https://huggingface.co/datasets/wltjr1007/Camelyon17-WILDS}{wltjr1007 mirror} and checked digests/metadata against the \href{https://github.com/p-lambda/wilds/blob/main/wilds/datasets/camelyon17_dataset.py}{official implementation}. Released centres 0/3/4 define the source; centres 1/2 supply evaluation only (Table~\ref{tab:data}). Source train/validation/internal-test patients are disjoint. These are hospital tests within CAMELYON, not separate releases. The source/derivative/mirror provenance is retained; CC0 centre identifiers have no authenticated institution-name mapping.
\subsection{Reference label} The target is \(y=1\) when the central 32\(\times\)32 window contains tumour; a negative patch can contain tumour outside it. Expert-pathologist annotations used cytokeratin immunohistochemistry for revision~\cite{camelyon}; isolated-tumour-cell annotation was non-exhaustive. This is dataset reference provenance, not clinician review of EF.
\begin{table*}[!t]\centering\small\setlength{\tabcolsep}{3pt}
\caption{CAMELYON17-WILDS data actually used. Released patient IDs are disjoint across these partitions. Only the first row updates model parameters.}
\label{tab:data}
\begin{tabular}{lrrrr}\toprule
Partition & Centres & Patches & Patients & Labels 0 / 1\\\midrule
Source training & 0, 3, 4 & 271,010 & 20 & 125,501 / 145,509\\
Source validation & 0, 3, 4 & 24,982 & 3 & 12,038 / 12,944\\
Internal evaluation & 0, 3, 4 & 40,004 & 3 & 30,459 / 9,545\\
External: CAMELYON centre 1 & 1 & 34,904 & 8 & 17,452 / 17,452\\
External: CAMELYON centre 2 & 2 & 85,054 & 9 & 42,527 / 42,527\\
\bottomrule\end{tabular}\end{table*}

\subsection{Quality controls} All patches were decoded and retained without caps. Shape/type, labels, patient/centre assignments, coordinates and decoded digests passed; no duplicate/conflict exclusion or patient overlap was recorded. Exact-byte separation does not establish independence of overlapping neighbourhoods. Scanner calibration, spacing and expert focus ratings are unavailable; no quality- or prediction-based exclusion was applied.
\begin{figure*}[!t]\centering
\includegraphics[width=\linewidth]{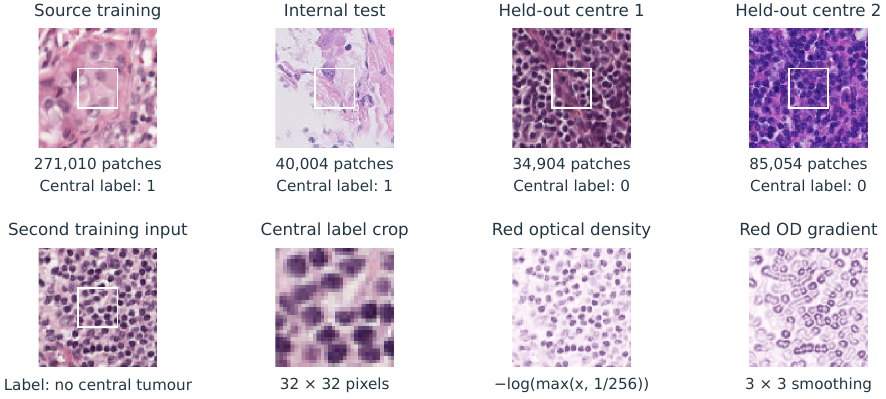}
\caption{Different source-training, internal-test and held-out-centre inputs (top). A second training input shows the central label crop and EF's red-channel optical-density and gradient measurements (bottom). White squares mark the released label window. The four top-row patients are disjoint; held-out-centre images never update model parameters. Optical views use display-only contrast stretching, and the crop is enlarged only for display.}\label{fig:views}\end{figure*}
\section{Evidence Field architecture}
\subsection{Optical observation contract} The input is \(x\in[0,1]^{3\times96\times96}\), obtained by exact uint8 RGB division by 255. EF retains three absorption channels rather than collapsing the image to a single intensity. It forms
\begin{equation}
\begin{aligned}
o_c(u)&=-\log\!\left(\max(x_c(u),1/256)\right),\\
q(u)&=\big[o(u),\sqrt{\lvert\nabla\bar o(u)\rvert^2+10^{-8}}\big]\in\mathbb R^6 .
\end{aligned}
\end{equation}
where \(\bar o\) is a 3\(\times\)3 valid-count average and derivatives use central differences with one-sided borders. The optical-density relation is established absorption physics~\cite{vahadane}; it does not by itself unmix stain concentrations or identify cells. Derivatives use pixel units because authenticated spacing is unavailable.
\subsection{Learned responses and same-image reference} Eight diagonal radial units share learned centres and positive precisions. Each mixes a source-standardized response with a current-image-standardized response:
\begin{equation}
\begin{aligned}
r_j(u)={}&\alpha_j e^{-\frac12\sum_c\lambda_{jc}(z_c(u)-a_{jc})^2}\\
&+(1-\alpha_j)e^{-\frac12\sum_c\lambda_{jc}(\tilde z_c(u)-a_{jc})^2}.
\end{aligned}
\label{eq:radial}
\end{equation}
Here \(\alpha_j=\sigma(\eta_j)\), \(\lambda_{jc}=\operatorname{softplus}(\ell_{jc})+10^{-3}\). Source normalization statistics are initialized from 32 uniformly selected training patches, independent of labels. The current image supplies its own channel mean and population standard deviation, floored at 0.03; a second source normalization acts on these relative coordinates. Coordinates are clipped to \([-6,6]\). Incoming-image statistics are computed without updating learned weights or using its label. The reference is not presumed healthy; robustness is assessed by centre transfer.
\subsection{Within-region structural signatures} Every region accumulates response mass, local centroid, covariance and a finite learned characteristic-function bank:
\begin{equation}
\begin{aligned}
\chi_{vj\ell}&=\frac{\sum_{u\in v}r_j(u)\exp\{i\omega_\ell^\top(u-c_v)\}}
{\max(\sum_{u\in v}r_j(u),10^{-6})},\\
\omega_\ell&=8\pi\tanh(\nu_\ell),\quad\ell=1,\ldots,4.
\end{aligned}
\label{eq:signature}
\end{equation}
\(c_v\) is the geometric midpoint in global normalized coordinates; centroid/covariance also use local half-width scaling. The implementation embeds 2D coordinates in three axes with the unused axis zero. Each response has ten mass/moment entries and eight real/imaginary signature entries. Concatenating own, parent-difference and root-difference descriptors with region mass, depth and a constant gives \(D=3\cdot8\cdot18+3=435\), followed by clipping. Arrangement-sensitive signatures augment low-order moments, but a finite bank cannot distinguish every possible arrangement.

Response and moment construction use FP64; completed descriptors and typed score maps use FP32; the hard covering composition uses FP64. The reconstruction contract is measured in Section~VII.
\subsection{Typed node and cross-scale scores} Let \(\mu_v\) be the pixel-area fraction of region \(v\). Two node types give \(a_k(v)=\mu_v\max_m\theta_{km}^\top\phi_v\). For a legal child pair \(G\), two interaction types augment the child anchors with four signed components:
\begin{align}
 \Pi_{kt}(G)&=\prod_{v\in G}\tanh(w_{kt}^{\top}\phi_v^{-}),\\
s_{kt}(G,p)&=\sum_{v\in G}a_k(v)+\mu_G\big[L_{kt}(G)+P_{kt}(p)\nonumber\\
&\qquad+\gamma_{kt}\Pi_{kt}(G)\nonumber\\
&\qquad+\gamma_{kt}\Pi_{kt}(G)\tanh(h_{kt}^{\top}\phi_p^{-})\big].
\label{eq:group}
\end{align}
The minus superscript omits the constant entry. \(L\) is an affine function of the mass-weighted child mean and absolute spread; \(P\) is an affine parent-context term. The remaining terms are a multiplicative child interaction and parent modulation. Descriptors already contain parent/root contrasts; the components describe computation rather than causal effects.
\subsection{Adaptive covering recursion} The depth-two spatial registry offers two alternative refinements per region. Each production partitions its parent; its legal pair packing \(\mathcal P\) leaves some children unmatched. Dynamic programming evaluates
\begin{align}
F_k(v)&=\max\{a_k(v),\mathcal C_k(v)\},\nonumber\\
\mathcal C_k(v)&=\max_{A,\mathcal P}
\bigg[\sum_{G\in\mathcal P}\max_t s_{kt}(G,v)\nonumber\\
&\hspace{48pt}+\sum_{u\in\operatorname{unmatched}(A,\mathcal P)}F_k(u)\bigg],\nonumber\\
z_k&=F_k(\mathrm{root})+b_k .
\label{eq:dp}
\end{align}
By induction on registry depth, a chosen node covers itself; a chosen production combines disjoint child supports. Every selected root program therefore covers the patch once. The recursion obtains the exact maximum over this finite legal registry, not over arbitrary anatomical segmentations. It records region bounds, production/packing, selected type, anchors, the four correction terms and bias. Soft log-sum-exp relaxations support training; hard inference and source validation use the declared maximum.
\subsection{Evidence identity} If \(\mathcal U_k\) is the selected program, \(z_k=b_k+\sum_{u\in\mathcal U_k}s_k(u)\). A signed display assigns each unit's score uniformly over its support, yielding \(\int e_k= z_k-b_k\). Thus \(\int(e_1-e_0)+(b_1-b_0)=z_1-z_0\). This display conserves classifier arithmetic; global references allow pixels outside a unit to affect its score.

\section{Training and evaluation protocol}
\subsection{Source partition and selection} All 271,010 training patches from centres 0, 3 and 4 enter every epoch. Validation uses 24,982 patches from three other source patients; 40,004 patches from three further source patients form the internal test. Centres 1 and 2 supply the two external hospital tests. Centre 1, used as out-of-distribution validation in the official WILDS protocol, is treated solely as evaluation here. Hence this study is not an exact official-leaderboard reproduction. The original training patient IDs are retained; label-blind alternating assignment of the remaining sorted source patient IDs gives the validation and internal-test partitions. All patient IDs and image ordering are frozen before training. Historical project use of these public targets makes the evaluation an exploratory replication; no target outcome selected the current objective, checkpoint, temperature or review threshold.
\subsection{Models and optimization} CNN is a from-scratch ResNet18~\cite{resnet}: a 7\(\times\)7 stride-two stem, four residual stages of widths 64/128/256/512 with two blocks each, global average pooling and a two-class linear readout. Our compact ViT~\cite{vit} partitions the image into 144 non-overlapping 8\(\times\)8 patches, projects each 192-value RGB patch to width 128, adds a class token and learned positions, and uses four pre-normalized blocks with eight attention heads, GELU, width-256 feed-forward units and dropout 0.1. Layer normalization and a two-class head read the class token. 
\subsection{Full-source fits} Each of six recipes is trained for 16 epochs with seeds 11, 29 and 47: 18 complete fits. AdamW uses learning rate 0.003 for EF and 0.001 for CNN/ViT, weight decay \(10^{-4}\), gradient norm clipping at 5, and cosine decay to \(10^{-5}\). Effective batch size is 64; EF accumulates two 32-image microbatches, while CNN/ViT use 64. Unweighted cross-entropy retains the actual source class distribution. A whole-patch quarter-turn/reflection is sampled for each effective batch; paired EF receives a further independently sampled geometric transform. Its label-preservation assumption is a registered stress assumption rather than a guarantee about all pathology transformations.
\subsection{Objectives and temperature} Single-view EF uses cross-entropy; paired EF averages cross-entropy over two transformed views. Probability-regularized EF adds \(0.05\) symmetric KL, and signed-evidence EF further adds \(0.02\) normalized aligned all-class spatial-evidence distance. All four share the same learned architecture and parameter count. Training relaxation is \(0.03\max(0,1-e/12)\), floored at 0.01 for paired recipes. Every epoch evaluates hard source logits. Checkpoints maximize source AUROC, breaking ties by lower stable cross-entropy and then earliest epoch. The EF headline recipe maximizes mean three-seed source AUROC, with mean cross-entropy as tie-break; it is paired cross-entropy. Four EF recipes versus one per comparator is an unequal family search budget, so single-view EF and all objectives are also reported.
\subsection{Prediction, uncertainty and review} An immutable 18-fit checkpoint lock precedes target inference. Each of 54 model/seed/pool prediction files is freshly reproduced and bound to image, patient, label, checkpoint and normalization digests. Headline probabilities are arithmetic means of three raw seed probabilities; calibrated ensembles average probabilities after each seed's source-only temperature. Temperature scaling~\cite{guo} minimizes source cross-entropy over 41 log-spaced values in \([0.25,4]\). Confidence cutoffs use source-validation quantiles for nominal 90/75/50/25\% retention. Evaluation reports actual external coverage and retained error without choosing an external cutoff. Percentile intervals use 2,000 patient-cluster bootstrap draws; paired differences share the same patient draws. All patches of each resampled patient remain together. External inference has eight and nine independent released patient groups.
\begin{table}[!t]\centering\small\setlength{\tabcolsep}{3pt}
\caption{Recorded full-source training cost per fit on A100 80 GB. Hours are mean and range over three seeds; memory is the maximum allocated device memory across seeds. EF paired CE is source selected.}
\begin{tabular}{lrrrr}\toprule
Recipe & Parameters & Mean h & Range h & Peak GiB\\\midrule
EF single-view & 10,550 & 1.14 & 1.13--1.16 & 0.847\\
EF paired CE & 10,550 & 1.96 & 1.88--2.03 & 1.362\\
CNN & 11,177,538 & 0.57 & 0.56--0.58 & 0.502\\
ViT & 573,826 & 0.48 & 0.47--0.49 & 0.363\\
\bottomrule\end{tabular}\par\vspace{3pt}
\begin{minipage}{\linewidth}\footnotesize Recorded time includes checkpointed complete epochs and the final successful attempt, with concurrent-worker contention; uncheckpointed failed-attempt overhead is not captured. Allocation includes architecture tensors and caches, but not all host/storage memory. Energy and isolated inference time for this current study are not measured here. A small parameter count does not imply faster training.\end{minipage}
\end{table}

\par

\section{Cross-centre classification and calibration}
\begin{table*}[!t]\centering\small\setlength{\tabcolsep}{3pt}
\caption{Full-set raw-probability ensembles. Brackets are 95\% AUROC patient-cluster percentile intervals (2,000 draws). Sensitivity and specificity refer to tumour as the positive class.}
\label{tab:results}
\begin{tabular}{llrrr}\toprule
Pool & Model & AUROC [95\% CI] & Accuracy \% & Sens. / spec. \%\\\midrule
Internal & EF paired CE & 0.9732 [0.856, 0.981] & 93.38 & 84.5 / 96.1\\
Internal & CNN & 0.9782 [0.895, 0.991] & 91.02 & 65.9 / 98.9\\
Internal & ViT & 0.9452 [0.786, 0.960] & 89.24 & 61.2 / 98.0\\
Centre 1 & EF paired CE & 0.9356 [0.831, 0.982] & 84.46 & 76.7 / 92.2\\
Centre 1 & CNN & 0.7352 [0.393, 0.869] & 58.94 & 25.3 / 92.6\\
Centre 1 & ViT & 0.8502 [0.587, 0.950] & 76.13 & 62.1 / 90.2\\
Centre 2 & EF paired CE & 0.9453 [0.905, 0.963] & 86.77 & 85.7 / 87.8\\
Centre 2 & CNN & 0.7656 [0.500, 0.845] & 67.60 & 44.5 / 90.7\\
Centre 2 & ViT & 0.7924 [0.632, 0.839] & 67.27 & 47.9 / 86.6\\
\bottomrule\end{tabular}\end{table*}

\begin{figure*}[!t]\centering\includegraphics[width=.90\textwidth]{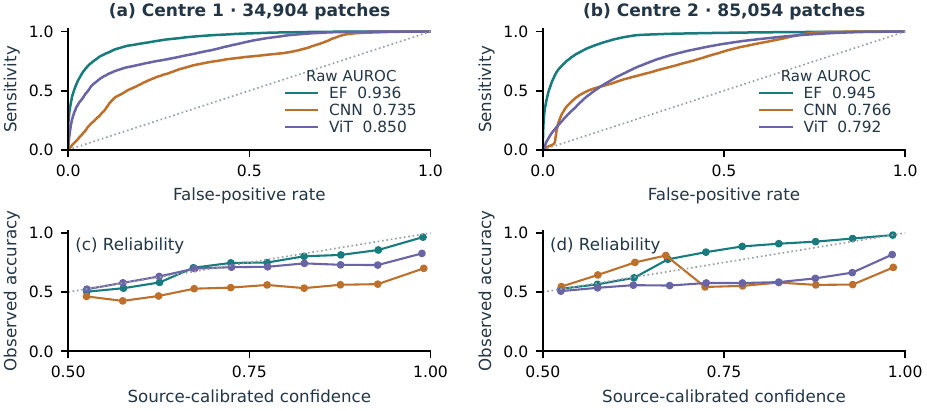}
\caption{External discrimination and confidence are separate results. Raw three-seed ROC curves and AUROC appear in (a,b); source-temperature reliability appears in (c,d). Every held-out patch enters its centre's curves. Reliability display bins use ten intervals from 0.5 to 1; tabulated ECE uses 15 fixed bins from 0 to 1. Dashed diagonals indicate random ranking and perfect calibration, respectively.}\label{fig:roc}\end{figure*}
\subsection{Paired patient uncertainty} Centre 1: EF--CNN AUROC difference [0.097, 0.444], EF--ViT AUROC difference [0.005, 0.252]. Centre 2: EF--CNN AUROC difference [0.108, 0.432], EF--ViT AUROC difference [0.114, 0.299]. These are exploratory percentile intervals, not multiplicity-adjusted superiority tests. There are 8 and 9 resampling units; all 2,000 AUROC draws were defined in each external comparison.

\begin{table*}[!t]\centering\small\setlength{\tabcolsep}{3pt}
\caption{All complete EF objective variants, each with 10,550 parameters. Source AUROC is a mean of three seed validation scores; test AUROC and accuracy use probability ensembles. Only source scores select the headline.}
\label{tab:objectives}
\begin{tabular}{lrrrrr}\toprule
EF objective & Source & Internal & Centre 1 & Centre 2 & External accuracy \%\\\midrule
EF single-view & 0.9894 & 0.9604 & 0.9416 & 0.9403 & 84.43 / 87.02\\
EF paired CE & 0.9905 & 0.9732 & 0.9356 & 0.9453 & 84.46 / 86.77\\
EF probability KL & 0.9899 & 0.9663 & 0.9423 & 0.9477 & 84.51 / 87.13\\
EF signed evidence & 0.9886 & 0.9608 & 0.9432 & 0.9451 & 84.23 / 87.06\\
\bottomrule\end{tabular}\end{table*}

\subsection{Scope of the comparison} EF leads the two implemented comparators in both external point estimates; CNN has the highest internal AUROC. Matching images and epochs leaves capacity, two-view exposure, wall time and family search effort unequal. EF objectives share capacity but are not node-only or parameter-matched architectural ablations. These measurements support the stated experiment rather than a universal clinical ranking.

\begin{table}[!t]\centering\small\setlength{\tabcolsep}{3pt}
\caption{Probability quality on all retained test images. ECE is shown before/after source-temperature calibration; Brier and NLL use calibrated ensembles.}
\label{tab:calibration}
\begin{tabular}{llrrr}\toprule
Pool & Model & ECE raw / cal. & Brier & NLL\\\midrule
Centre 1 & EF paired CE & 0.027 / 0.030 & 0.109 & 0.350\\
Centre 1 & CNN & 0.281 / 0.247 & 0.297 & 1.192\\
Centre 1 & ViT & 0.138 / 0.125 & 0.193 & 0.717\\
Centre 2 & EF paired CE & 0.037 / 0.031 & 0.095 & 0.311\\
Centre 2 & CNN & 0.235 / 0.210 & 0.270 & 0.921\\
Centre 2 & ViT & 0.191 / 0.170 & 0.239 & 0.778\\
\bottomrule\end{tabular}\end{table}

\section{Outputs beyond a classifier probability}
\subsection{Measured independent evidence} A fixed panel takes the first ten released-order patches in each of 16 patient/label strata across eight centre-1 patients: 160 distinct real images, balanced across labels. Seeds 11 and 29 supply 320 image--checkpoint pairs; original, quarter-turn and reflection inputs yield 960 transformed records and 1920 class programs. Fresh source-checkpoint coefficient comparison and a separate framework-free reader passed. Maximum signed decomposition error was 8.76e-06; maximum root-optimum error 8.76e-06; maximum associated prediction-logit error 5.96e-06. The mean/median aligned signed distance was 0.176/0.168, and mean typed-cover disagreement 0.878. For the 320 original records, descriptor error was at most 1.19e-07; all 2,834 reachable routes matched the independent canonical choice, and each class covered exactly 9,216 pixels. This is a fixed mathematical audit panel, not a diagnostic-accuracy sample or an all-test-patch native reconstruction.

\par
\begin{figure*}[!t]\centering\includegraphics[width=\linewidth]{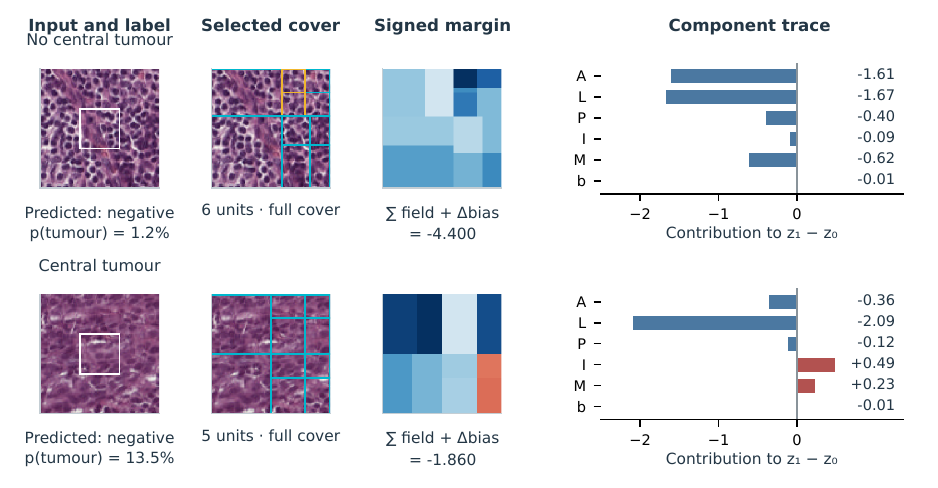}
\caption{Two unchanged centre-1 examples from the fixed audit panel. Raw seed-11 EF outputs include the prediction, selected class-1 node/group supports (yellow/cyan), signed class-difference field and term accounting. A/L/P/I/M/b denote anchors, local affine, parent affine, member interaction, parent modulation and bias; their signed differences sum to the logit margin. Red/blue mean positive/negative contributions; field scales are panel-specific. The lower tumour-positive example is missed. These computational supports do not establish lesion localization.}\label{fig:evidence}\end{figure*}
\noindent Exact component totals accompanying Figure~\ref{fig:evidence} are retained in the public numerical audit record.
\par
\subsection{Additional outputs and verification} EF supplies structural signatures, selected scale/refinement and node/pair types, five signed non-bias totals, a once-only cover and an authenticated reconstruction target. Its independent reader regenerates descriptors, typed scores, reachable routes and logits from the input and source-checkpoint coefficients. These CNN/ViT implementations do not emit covering-program records; other explanation methods are separate comparators.
\subsection{Aligned stability} The signed-evidence training variant uses the
uncentered distance \(\|e-e'\|_1/(\|e\|_1+\|e'\|_1+10^{-12})\), also used for
paired-CE panel diagnostics. The headline paired-CE model has no evidence-consistency term.
Typed-cover disagreement is the fraction of aligned pixel/class entries whose
support-union identifier, node/group kind or selected type differs. The mean
0.878 is a program-disagreement measurement, not a localization or stability
guarantee. The separately completed comparison in the related architecture study~\cite{efcore} trained three class-centred-loss fits under the same source-locked protocol. It gives held-out areas under the curve of 0.9420 and 0.9494, compared with 0.9432 and 0.9451 for the original signed variant. Paired patient-bootstrap differences include zero at both centres. Those fits do not replace the primary paired-cross-entropy ensemble or retrospectively alter its training loss. Variant-specific aligned-stability and full-native-replay measurements remain unavailable; matched paired-view baseline fits were omitted from the shortened validation study.

\subsection{Ensemble replay} Figure~\ref{fig:evidence} explains seed 11.
An expanded panel retains the first ten released-order patches per patient/label
stratum: 160 centre-1 and 180 centre-2 images, covering all 17 released patients.
A fresh NumPy/SciPy reader reconstructs all three members under three rigid views
(3,060 member records; 6,120 class programs), verifying descriptors, types,
reachable routes, one-fold covers, signed totals and optima. Base inputs and
logits are bound to the full-set saved predictions. Source temperatures $T_j>0$
and weights $w_j=1/3\geq0$, $\sum_jw_j=1$, reproduce raw and calibrated mixtures
with maximum error \(2.81\times10^{-7}\); maximum descriptor and optimum errors
are \(1.19\times10^{-7}\) and \(6.38\times10^{-6}\).
This fixed panel verifies computation; classification estimates use all released test patches.
A mean logit or one member's field does not reconstruct the ensemble.
\begin{table}[!t]\centering\small\setlength{\tabcolsep}{3pt}
\caption{EF source-fixed review policies on both external centres. Retained error is conditional on accepted patches; brackets are 95\% patient-cluster intervals. Source retention targets 75/90\% use confidence cutoffs 0.9564/0.8253. Reported retention is the achieved external coverage.}
\label{tab:review}
\begin{tabular}{lrrrr}\toprule
Pool & Policy \% & Retain. \% & Error \% [95\% CI]\\\midrule
Centre 1 & 75 & 51.79 & 3.37 [0.76, 9.01]\\
Centre 1 & 90 & 68.84 & 6.46 [1.71, 15.24]\\
Centre 2 & 75 & 34.91 & 1.64 [1.19, 2.65]\\
Centre 2 & 90 & 60.90 & 3.53 [2.31, 5.90]\\
\bottomrule\end{tabular}\end{table}

\subsection{Pathology review output} An incoming patch returns class probabilities, source-calibrated confidence, region bounds/types, signed components and a checkpoint-bound program. Figure~\ref{fig:evidence} shows the central target and computed margin. Institutions can check provenance and replay, then apply a locally validated review policy. Computational reconstruction is distinct from diagnostic correctness; clinical usefulness and time savings require prospective evaluation.
\section{Conclusion and reproducibility}
EF combines learned structure with signed spatial classification. Its source-selected ensemble, with 10,550 parameters per member, exceeds the implemented CNN/ViT on both held-out centres. Patient uncertainty, calibration, costs and exploratory public-target history define that result's scope. The reproducibility record retains preparation, 18 checkpoints/histories, source policies, 54 prediction receipts, statistics, frozen source and independently reconstructed programs. Code, the independent reader, configurations and a reproducible input are available at \url{https://github.com/isararsiddique/evidence-field}; model states and hash-bound artifacts are available at \url{https://huggingface.co/Isarar/Evidence-Field}.

\section*{Acknowledgment}
Isarar Siddique conceived and planned the study, led the research, and coordinated the project. Raiyan Siddique, Arshi Niyaz, Jamyang Ugyen Tshomo, and Atish Raza assisted with the research. The authors are responsible for the scientific content, verification, and final submission.

OpenAI Codex assisted with drafting and revision (Abstract and Sections I--VIII), exposition of the study protocol and mathematical model, figure-layout code (Figs.~1--4), and audit scripts. Input images and reported numerical findings are retained from the declared public release and bound study records.

The authors declare that they have no conflicts of interest.

\IEEEtriggeratref{30}

\end{document}